\documentclass[journal]{IEEEtran}
\def\baselinestretch{1.2}

\usepackage{tikz}
\graphicspath{{images/}}
\usepackage{amsmath}
\usepackage{latexsym, amssymb}
\usepackage{graphicx}
\usepackage{amsmath, amsbsy}
\usepackage{amsopn, amstext}
\usepackage{ifpdf,hyperref}
\usepackage{cancel, color}
\usepackage{epstopdf}

\def\lvtimes{\vec{\ltimes}}
\def\rvtimes{\vec{\rtimes}}

\def\J{{\bf 1}}

\DeclareMathOperator{\Col}{Col}

\DeclareMathOperator{\lcm}{lcm}

\def\cal{\mathcal}

\def\ra{\rightarrow}

\def\lra{\leftrightarrow}

\def\a{\alpha}
\def\b{\beta}

\def\0{{\bf 0}}

\def\argmin{argmin}

\def\spark{spark}
\newcommand{\R}{{\mathbb R}}

\newcommand{\Z}{{\mathbb Z}}

\def\dsum{\mathop{\sum}\limits}

\newtheorem{thm}{Theorem}[section]
\newtheorem{dfn}[thm]{Definition}
\newtheorem{prp}[thm]{Proposition}

\newtheorem{cor}[thm]{Corollary}
\newtheorem{rem}[thm]{Remark}

\begin{document}

\title{ Signal Processing via Cross-Dimensional Projection}
\author{Daizhan Cheng
	\thanks{This work is supported partly by the National Natural Science Foundation of China (NSFC) under Grants 62073315.}
\thanks{Key Laboratory of Systems and Control, Academy of Mathematics and Systems Sciences, Chinese Academy of Sciences,
		Beijing 100190, P. R. China (e-mail: dcheng@iss.ac.cn).}
}

\maketitle

\begin{abstract}
Using projection between Euclidian spaces of different dimensions, the signal compression and decompression become  straightforward. This encoding/decoding technique requires no preassigned measuring matrix as in compressed sensing. Moreover, in application there is no dimension or size restrictions. General formulas for encoding/decoding of any finite dimensional signals are provided. Their main properties are revealed. Particularly, it is shown that under the equivalence assumption the technique provides the best approximation with least square error.
\end{abstract}


\IEEEpeerreviewmaketitle

\section{Introduction}

Signal compression-decompression is a fundamental task for signal processing. Various methods have been developed. to deal with this problem. For instance, in image compression-decompression, the discrete fourier transformation, wavelet transformation, discrete cosine transformation, etc. have been used \cite{pen93}.  The basic idea in handling signal compression-decompression is like this. Using a method to transfer a signal, which is digitalized as a point  $x\in \R^n$ space, to a point  $y\in \R^m$ space, where $m<n$. This process is called signal compression. Then use another transformation to convert $y$ back to $
\hat{x}\in |R^n$. This process is called signal decompression. To judge the quality of a complression/decompression method, usually we use the square error \cite{dua11}
\begin{align}\label{1.1}
Error=\|\hat{x}-x\|_2=\sqrt{\Sigma_{i=1}^n(\hat{x}_i-x_i)^2}.
\end{align}

In all existing compression-decompression methods, since $y$ is of completely different kind of measurements, the analysis of error becomes difficult.

This paper proposes a new compression-decompression method, which makes the mixed dimensional Euclidian spaces
$$
\R^{\infty}=\bigcup_{n=1}^{\infty}\R^n
$$
into a ``vector space". Using triangular inequality, we have
\begin{align}\label{1.2}
Error\leq \|\hat{x}-y\|_2+\|y-x\|_2.
\end{align}
Then we try to minimize both $\|\hat{x}-y\|_2$ and $\|y-x\|_2$ to get best recovered signal.
\begin{align}\label{1.3}
\min_{y\in \R^m}\|y-x\|_2
\end{align}
yields the best compressed signal, and
\begin{align}\label{1.4}
\min_{\hat{x}\in \R^n}\|y-\hat{x}\|_2
\end{align}
yields the best decompressed (recovered) signal.

After establishing a metric vector space structure on $\R^{\infty}$, both compression and decompression can be executed by proper projections. This is the basic idea of this paper.

\section{Preliminaries}

This section  reviews semi-tensor product (STP) of matrices, the vector space structure and topological structure on $\R^{\infty}$. We refer to \cite{che12,che16} for more details.

\subsection{Semi-Tensor Product of Matrices}

\begin{dfn}\label{d2.0.1}
\begin{itemize}
\item[(i)] Assume $A\in {\cal M}_{m\times n}$,   $B\in {\cal M}_{p\times q}$, and $\lcm(n,p)=t$. Then the left matrix-matrix (MM-) STP of $A$ and $B$ is defined as
\begin{align}\label{2.0.1}
A\ltimes B:=\left(A\otimes I_{t/n}\right) \left(B\otimes I_{t/p}\right)\in {\cal M}_{mt/n\times qt/p}.
\end{align}
\item[(ii)] Assume $A,B$ are as in (i). Then the right (MM-) STP of $A$ and $B$ is defined as
\begin{align}\label{2.0.2}
A\rtimes B:=\left( I_{t/n}\otimes A\right) \left(I_{t/p}\otimes B\right)\in {\cal M}_{mt/n\times qt/p}.
\end{align}
\item[(iii)] Assume $A\in {\cal M}_{m\times n}$,   $x\in \R^p$,  and $\lcm(n,p)=t$. Then the left matrix-vector (MV-) STP of $A$ and $x$ is defined as
\begin{align}\label{2.0.3}
A\lvtimes x:=\left(A\otimes I_{t/n}\right) \left(x\otimes \J_{t/p}\right)\in R^t.
\end{align}
\item[(iv)] Assume $A$ and $x$ are as in (iii).  Then the right matrix-vector (MV-) STP of $A$ and $x$ is defined as
\begin{align}\label{2.0.4}
A\rvtimes x:=\left(I_{t/n}\otimes A\right) \left(\J_{t/p}\otimes x\right)\in R^t.
\end{align}
\end{itemize}
\end{dfn}

The basic properties of STPs are listed as follows:

\begin{prp}\label{p2.0.2}\cite{che12} Let $A,B,C\in {\cal M}$, $x,y\in \R^{\infty}$.
\begin{enumerate}
\item[(1)] (Consistency)
\begin{itemize}
\item[(i)] When $n=p$, the MM-STPs are degenerated to conventional MM-product. That is,
\begin{align}\label{2.0.5}
A\ltimes B=A\rtimes B=AB.
\end{align}
\item[(ii)] When $n=r$, the MV-STPs are degenerated to conventional MV-product. That is,
\begin{align}\label{2.0.6}
A\lvtimes x=A\rvtimes x=Ax.
\end{align}
\end{itemize}
\item[(2)] (Associativity)
\begin{itemize}
\item[(i)]
\begin{align}\label{2.0.7}
(A\ltimes B)\ltimes C=A\ltimes (B\ltimes C).
\end{align}
\item[(ii)]
\begin{align}\label{2.0.8}
(A\rtimes B)\rtimes C=A\rtimes (B\rtimes C).
\end{align}
\item[(iii)]
\begin{align}\label{2.0.9}
(A\ltimes B)\lvtimes x:=A\lvtimes (B\lvtimes x).
\end{align}
\item[(iv)]
\begin{align}\label{2.0.10}
(A\rtimes B)\rvtimes x:=A\rvtimes (B\rvtimes x).
\end{align}
\end{itemize}
\item[(3)] (Distributivity)
\begin{itemize}
\item[(i)]
\begin{align}\label{2.0.11}
\begin{array}{l}
(A + B)\ltimes C=A\ltimes C+B\ltimes C,\\
C\ltimes (A + B)=C\ltimes A+C\ltimes B.
\end{array}
\end{align}
\item[(ii)]
\begin{align}\label{2.0.12}
\begin{array}{l}
(A + B)\rtimes C=A\rtimes C+B\rtimes C,\\
C\rtimes (A + B)=C\rtimes A+C\rtimes B.
\end{array}
\end{align}
\item[(iii)]
\begin{align}\label{2.0.13}
\begin{array}{l}
(A+B)\lvtimes x=A\lvtimes x+B\lvtimes x,\\
A\lvtimes (x+y)=A\lvtimes x+A\lvtimes y.\\
\end{array}
\end{align}
\item[(iv)]
\begin{align}\label{2.0.14}
\begin{array}{l}
(A+B)\rvtimes x=A\rvtimes x+B\rvtimes x,\\
A\rvtimes (x+y)=A\rvtimes x+A\rvtimes y.\\
\end{array}
\end{align}
\end{itemize}
\end{enumerate}
\end{prp}

\subsection{Vector Space Structure on $\R^{\infty}$}

\begin{dfn}\label{d2.1.1}  Let $x,y\in \R^{\infty}$ with $x\in \R^m$ and $y\in \R^n$, $t=\lcm(m, n)$.
\begin{itemize}
\item[(i)] The addition/subtraction on $\R^{\infty}$ is defined  by
\begin{align}\label{2.1.1}
x\pm_{\ell} y:=\left(x\otimes \J_{t/m}\right)\left(y\otimes \J_{t/n}\right)\in \R^t.
\end{align}
or
\begin{align}\label{2.1.2}
x\pm_{r} y:=\left(\J_{t/m}\otimes x\right)\left(\J_{t/n}\otimes y\right)\in \R^t.
\end{align}
\item[(ii)] The scalar product of $x$ is as a conventional scalar product, i.e.,
\begin{align}\label{2.1.3}
r\times x:=rx\in  \R^m,\quad r\in \R.
\end{align}
\end{itemize}
\end{dfn}

%\begin{rem}\label{r2.1.2} Both left addition/subtraction (\ref{2.1.1}) and right addition/subtraction (\ref{2.1.2}) have been defined in \cite{che16}. From the perspective of signal processing but most previous discussions are concerned on left ones. It is straightforward to pass the results about left addition/subtraction to right ones. In the following discussion the inner product, norm, distance, etc. have been defined and discussed for left ones. We briefly call them the left system. They can all be easily passed to right system. Since in signal process the right system is more suitable in use, we will directly state the known results for left system as the corresponding results for right systems without further explanation.
%
%Hereafter, we consider the right system as the default one. That is, briefly set $\pm：=\pm_{r}$.
%\end{rem}

\begin{prp}\label{p2.1.2} With the addition defined by (\ref{2.1.1}) or (\ref{2.1.2}) and scalar product defined by (\ref{2.1.3}), the $\R^{\infty}$ becomes a pseudo-vector space.\footnote{A pseudo-vector space satisfies all the requirements for a vector space except that $x-y=0$ does not implies $x=y$ \cite{abr78}}.
\end{prp}

\begin{dfn}\label{d2.1.3} Two vectors $x$ and $y$ are said to be left (right) equivalent, denoted by $x\lra_{\ell} y$ ($x\lra_{r} y$), if $x -_{\ell}y=0$ ($x-_{r}y=0$).
The equivalence class of $x$ is denoted by $\bar{x}_{\ell}$. ($\bar{x}_{r}$). That is,
$$
\bar{x}_{\ell}:=\{y\in \R^{\infty}\;|\;y\lra_{\ell} x\}, \quad (\bar{x}_{r}:=\{y\in \R^{\infty}\;|\;y\lra_{r} x\}).
$$
\end{dfn}

\begin{rem}\label{r2.1.4} From signal point of view, the left equivalence of sampling data can be understood as the same signal with different sampling frequencies, while the right equivalence can be understood as the same signal with different sampling periods.
Numerical experience shows that in most cases the left system seems more efficient than the right system. Hence to make arguments simple, hereafter, we will concentrate on ``left system" and consider the left objects as the default ones.  That is, assume
 \begin{align}\label{2.1.301}
\pm:=\pm_{\ell};\quad \lra:=\lra_{\ell};\quad \bar{x}:=\bar{x}_{\ell}; \quad \mbox{etc.}
\end{align}
\end{rem}

\begin{prp}\label{p2.1.5} Let $x,y\in \R^{\infty}$.
\begin{itemize}
\item[(i)]  $x\lra y$, if and only if, there exist two one-vectors $\J_{\a}$ and $\J_{\b}$, such that
 \begin{align}\label{2.1.4}
x\otimes \J_{\a}=y\otimes \J_{\b}.
\end{align}
\item[(ii)] If $x\lra y$, then there exists a $z\in \R^{\infty}$ such that
 \begin{align}\label{2.1.5}
x=z\otimes \J_{b};\quad y=z\otimes \J_{a}.
\end{align}
\item[(iii)] Without loss of generality, we assume in (\ref{2.1.4}) $\a$ and $\b$ are co-prime, i.e., $\gcd(\a,\b)=1$. Then in (\ref{2.1.5})
$$
a=\a;\quad b=\b.
$$
\end{itemize}
\end{prp}

%Note that from previous proposition one sees easily that the equivalence of two signals has clear physical meaning. Since (ref{2.1.5}) means
%$$
%\begin{array}{l}
%x=(\underbrace{z^T,z^T,\cdots,z^T}_b)^T,\\
%y=(\underbrace{z^T,z^T,\cdots,z^T}_a)^T,\\
%\end{array}
%$$
%$x\lra y$ means $x$ and $y$ are generated from the same signal $z$ but with different sampling lengths. So from frequency domain,
%they are exactly the same.

\subsection{Topological Structure on $\R^{\infty}$}

\begin{dfn}\label{d2.2.1}  Let $x,y\in \R^{\infty}$ with $x\in \R^m$ and $y\in \R^n$, $t=\lcm(m, n)$.
\begin{itemize}
\item[(i)]
The inner product of $x$ and $y$ is defined by
\begin{align}\label{2.2.1}
\langle x,y\rangle_{{\cal V}}:=\frac{1}{t} \langle (x\otimes \J_{t/m}),(y\otimes \J_{t/n}) \rangle,
\end{align}
where
$$
\langle (x\otimes \J_{t/m}),(y\otimes \J_{t/n}) \rangle=(x\otimes \J_{t/m})^T(y\otimes \J_{t/n})
$$
is the conventional inner product on $\R^t$.

\item[(ii)] The norm of $x\in \R^n$ is defined by
\begin{align}\label{2.2.2}
\|x\|_{{\cal V}}:=\sqrt{\langle x,x\rangle_{{\cal V}}}=\frac{1}{\sqrt{n}}\sqrt{x^Tx}.
\end{align}
\item[(iii)] The distance of $x$ and $y$ is defined by
\begin{align}\label{2.2.3}
d_{{\cal V}}(x,y):= \|x\vec{-}y\|_{{\cal V}}.
\end{align}
\end{itemize}
\end{dfn}

\begin{rem}\label{r2.2.101} It is easy to verify that
if $x,y\in \R^n$ then
\begin{align}\label{2.2.301}
d_{{\cal V}}(x,y)=\frac{1}{\sqrt{n}}d_2(x,y).
\end{align}
Hence when the $d_{{\cal V}}$ is restricted on a fixed dimensional Euclidian space, it is proportional to standard distance.
\end{rem}

Using this inner product, the angle between two elements  $x,y\in \R^{\infty}$, denoted by $\theta=\theta_{x,y}$, is defined by
\begin{align}\label{2.2.4}
\cos(\theta)=\frac{\langle x,y\rangle_{{\cal V}}}{\|x\|_{{\cal V}}\|y\|_{{\cal V}}}.
\end{align}

Under this distance, the distance deduced topology can be obtained, which is denoted by ${\cal T}_d$. Apart from this ${\cal T}_d$, there is another topology on $\R^{\infty}$. Naturally, each $\R^n$ can be considered as a component (i.e., a clopen set) of $\R^{\infty}$. And within each $\R^n$ the conventional Euclidian space topology is used. Then overall, such a topology is called the natural topology, denoted by ${\cal T}_n$. $(\R^{\infty},{\cal T}_n)$ is disconnected. Precisely speaking, its fundamental group is $\Z^+$.

It is easy to verify the  following result.

\begin{prp}\label{p2.2.2}Consider $\R^{\infty}$. Let $x,y\in \R^{\infty}$. Then
$d(x,y)=0$, if and only if, $x\lra y$.
\end{prp}

\begin{prp}\label{p2.2.3}  The inner product defined by (\ref{3.2.1}) is consistent with the equivalence. That is,
if $x_1\lra x_2$ and $y_1\lra y_2$, then
\begin{align}\label{2.2.5}
\langle x_1,y_1\rangle_{{\cal V}}=\langle x_2,y_2\rangle_{{\cal V}}.
\end{align}
\end{prp}

According to Proposition \ref{p2.2.3}, the norm and distance defined by Definition \ref{d2.2.1} for $\R^{\infty}$ are also consistent with the equivalence.

In fact, under the equivalence (or the distance) the space of finite dimensional signals form a quotient space
\begin{align}\label{2.2.6}
\Omega:=\R^{\infty}/\lra.
\end{align}
We call $\Omega$ the signal space. Then this signal space has the following properties.

\begin{prp}\label{p2.3.4} Consider the signal space $\Omega$.
\begin{itemize}
\item[(i)] It is a topological space with the quotient topology
$$
{\cal T}_d={\cal T}_n/\lra.
$$
\item[(ii)] As a topological space $(\Omega,{\cal T}_d)$ is second countable, Hausdorff, separable. (We refer to any text book on topology for the related concepts.)
\item[(iii)] The addition (\ref{2.1.1}) and scalar product (\ref{2.1.3}) can be extended to $\Omega$ as
\begin{align}\label{2.2.7}
\begin{array}{l}
\bar{x}\pm \bar{y}:=\overline{x\pm y},\\
r\bar{x}:=\overline{rx}.
\end{array}
\end{align}
Then with the addition and scalar product defined in (\ref{2.2.7}), $\Omega$ becomes a vector space. Hence $\Omega$ is a topological vector space. (We refer to \cite{kel63} for related concepts and properties.)
\end{itemize}
\end{prp}

\subsection{Projection between Two Euclidian Spaces with Different Dimensions}

\begin{dfn}\label{d2.3.1} Let $x\in \R^m$. The projection of $x$ onto $\R^n$, denoted by $\pi^m_n(x)$, is defined by
\begin{align}\label{2.3.1}
\pi^m_n(x)=\argmin_{y}(d_{{\cal V}}(y,x)).
\end{align}
\end{dfn}

\begin{prp}\label{p2.3.2} \cite{che16,che19}
\begin{itemize}
\item[(i)] The projection of $x\in \R^m$ onto $\R^n$ is
\begin{align}\label{2.3.2}
y_0:=\pi^m_n(x)=\Pi^m_n\rtimes x,
\end{align}
where ($t=\lcm(m,n)$)
$$
\Pi^m_n=\frac{n}{t}\left(I_n\otimes \J^T_{t/n}\right)\left(I_m\otimes \J_{t/m}\right).
%\Pi^m_n=\frac{n}{t}\left(\J^T_{t/n}\otimes I_n\right)\left(\J_{t/m}\otimes I_m\right).
$$
\item[(ii)] $x\vec{-} y_0$ is perpendicular to $y_0$,  denoted by $(x\vec{-} y_0)\perp y_0$.  That is, the angle between them, denoted by $\theta$ satisfies
\begin{align}\label{2.3.3}
\cos(\theta)=0.
\end{align}
\item[(iii)] For any $y\in \R^n$, $(x\vec{-} y_0)\perp y$.
\end{itemize}
\end{prp}

\begin{figure}
\centering
\setlength{\unitlength}{6mm}
\begin{picture}(7,4)\thicklines
\put(3,1){\vector(1,0){3}}
\put(3,1){\vector(1,1){3}}
\put(6,1){\vector(0,1){3}}
\put(6,1){\vector(-3,-1){2}}
\put(2,1.4){$\R^n$}
\put(4.5,3){$x$}
\put(6.2,2.5){$x\vec{-} y_0$}
\put(4.5,1.2){$y_0$}
\put(3.7,0.3){$y$}
\put(2.6,0.8){$0$}
\thinlines
\put(0,0){\line(1,0){6}}
\put(2,2){\line(1,0){6}}
\put(0,0){\line(1,1){2}}
\put(6,0){\line(1,1){2}}

\end{picture}
\caption{Projection \label{Fig2.1}}
\end{figure}
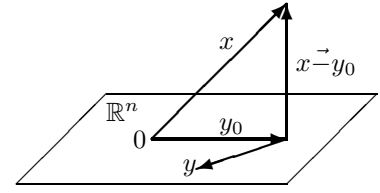

(See Figure \ref{Fig2.1}.)

\section{Projection-based Signal Compression/Decompression}

This section consider the projection-based compression (PC) and the projection-based decompression (PD). First, the one dimensional signals are considered and then the higher dimensional signals are considered.

\subsection{PS and PD for $x\in \R^{\infty}$}

Assume $x\in \R^{\infty}$ is a finite dimensional signal, specified by $x\in \R^n$. We want to compress it into
$y\in \R^m$ ($m<n$). And then recover it back as $\hat{x}\in \R^n$.

\begin{dfn}\label{d3.1.1}
\begin{itemize}
\item[(i)] The PS of $x\in \R^n$ to $y\in \R^m$ is defined by
\begin{align}\label{3.1.1}
y=\pi^n_m(x)=\Pi^n_mx.
\end{align}
\item[(ii)] The PD of $y\in \R^m$ to $\hat{x}\in \R^n$ is defined by
\begin{align}\label{3.1.2}
\hat{x}=\pi^m_n(y)=\Pi^m_n y.
\end{align}
\end{itemize}
\end{dfn}

\begin{thm}\label{t3.1.1} Under the assumption that $x\lra y$ means they are the same. (The physical meaning of the equivalence has been explained before.) PR and PD form the best compression-decompression pair.
\end{thm}

\noindent{\it Proof.} Any signal compression-decompression can be described by two mappings:
 the compressing mapping $\phi:\R^n\ra \R^m$ and the decompressing mapping $\psi:\R^m\ra \R^n$.

 Under the assumption that $x\lra y$ implies they are the same,  we know that
\begin{align}\label{3.1.3}
x=x\otimes \J_{\a},\quad \forall \a\in \Z^+.
\end{align}

Consider any ``distance", denoted by $d_*$ , which measures the fixing degree of signal $x$ with its compressed measurement $y=\phi(x)$  as $d_*(x,y)$. Note that $d_*$ needs not to be a real distance,  it is proper as long as it can measure the quality of the compression. Say, we can define
\begin{align}\label{3.1.4}
d_*(x,y):=d_n(\psi(y),x),
\end{align}
where $d_n$ is the standard distance on $\R^n$.

Assume $t=\lcm(m,n)$ and denote $\tilde{x}=x\otimes \J_{t/n}$ and $\tilde{y}=y\otimes \J_{t/m}$. Under the assumption (\ref{3.1.3}) we have
$$
d_*(x,y)=d_*(\tilde{x}, \tilde{y}).
$$

Now $\tilde{x}$ and $\tilde{y}$ lie on the same dimensional Euclidian space $\R^t$. Since a common measurement of two vectors (in a same dimensional Euclidian space) is  $\ell_2$ norm as in (\label{1.1}), it is reasonable to assume that when restrict on each $\R^t$, $d_*$ is order-preserving  with $\|\cdots\|_2$, which means when $u,v,u',v'\in \R^t$, $d_*(u,v)>d_*(u',v')$, if and only if,$\|u-v\|_2>\|u'-v'\|_2$. (Note that this argument is based on the fact that $\ell_2$ norm is commonly used to measure the error \cite{dua11}. Hence, taking Remark \ref{r2.2.101} into consideration, we have
 \begin{align}\label{3.1.5}
\begin{array}{l}
\min_{y\in \R^m} d_*(y,x)
=\min_{y\in \R^m}d_*(\tilde{x},\tilde{y})\\
=\min_{y\in \R^m}d_n(\tilde{x},\tilde{y})\\
=\min_{y\in \R^m}d_{{\cal V}}(\tilde{x},\tilde{y})\\
=\min_{y\in \R^m}d_{{\cal V}}(x,y)\\
\end{array}
\end{align}

Now by definition of the projection, it is clear that
$$
y=\pi^n_m(x)
$$
is the unique best solution to (\ref{3.1.5}).

Similarly
$$
\hat{x}=\pi^m_n(y)
$$
is the best solution for the decomposition.

\hfill $\Box$

\subsection{PS and PD for Higher Dimensional Signals}

To deal with $n\geq 2$ dimensional signals we need to express the signals into matrices or hypermatrices.

\begin{dfn}\label{d3.2.1} A hypermatrix with order $d$ is a set of ordered data, which can be expressed by
\begin{align}\label{3.2.1}
A=\{a_{i_1,i_2,\cdots,i_d}\;|\; i_k\in [1,n_k],~k=1,2,\cdots,d\}.
\end{align}
The set of hypermatrices with order $d$ and ranges $n_1,n_2,\cdots,n_d$ is denoted by
$$
\R^{n_1\times \cdots\times n_d}.
$$
\end{dfn}

A matrix can be considered as a hypermatrix with order $2$. So the following definition is also applicable to matrices.

\begin{dfn}\label{d3.2.2} \cite{lim13} Let $A=(a_{i_1,\cdots,i_p})\in \R^{n_1\times \cdots \times n_p}$ and $B=(b_{j_1,\cdots,j_q})\in \R^{m_1\times \cdots\times m_q}$.
Assume $i_s$ and $j_t$ have the same range, i.e., $n_s=m_t$, then the contraction product of $A$ and $B$, denoted by
\begin{align}\label{3.2.2}
C=A@^s_t B,
\end{align}
where
$$
C=\{c_{i_1\cdots \hat{i}_s,\cdots,i_p,j_1,\cdots,\hat{j}_t,\cdots,j_q}\},
$$
satisfies
\begin{align}\label{3.2.3}
c_{i_1\cdots \hat{i}_s,\cdots,i_p,j_1,\cdots,\hat{j}_t,\cdots,j_q}:=\dsum_{k=1}^{n_s} a_{i_1\cdots, k,\cdots,i_p}b_{j_1,\cdots,k,\cdots,j_q},
\end{align}
\end{dfn}
where by convention a caret over any entry means that the respective entry is omitted.

Note that when the order is $2$ the matrix product of $A=(a_{i,j})\in {\cal M}_{m\times n}$ and $B=(b_{p,q})\in {\cal M}_{n\times t}$ can be considered as
$$
AB=A@^j_p B.
$$
Hence, Definition \ref{d3.2.2} coincides with the conventional matrix product.

An image is a finite dimensional signal with order $2$, which can be expressed as a matrix $A\in {\cal M}_{m\times n}\subset {\cal M}$.
A stereopicture is a finite dimensional signal with order $3$. A finite dimensional signal with its order $n\geq 3$ can be expressed by a hypermatrix of order $n$.

\begin{prp}\label{p3.2.3} Let $A\in \R^{n_1\times \cdots \times n_d}$ be a $d$-dimensional signal, where $d\geq 2$.
\begin{itemize}
\item[(i)] The projection-based compression of $A$ into $B\in \R^{m_1\times\cdots \times m_d}$ can be realized by
\begin{align}\label{3.2.4}
B:=\left(\Pi^{n_1}_{m_1}@^2_{i_1} \right)\left(\Pi^{n_2}_{m_2}@^2_{i_2} \right)\cdots \left(\Pi^{n_d}_{m_d}@^2_{i_d} \right)
A.
\end{align}
\item[(ii)] The projection-based decompression of $B\in \R^{m_1\times \cdots \times m_d}$ into $\hat{A}$ can be realized by
\begin{align}\label{3.2.5}
\hat{A}:=\left(\Pi^{m_1}_{n_1}@^2_{i_1} \right)\left(\Pi^{m_2}_{n_2}@^2_{i_2} \right)\cdots \left(\Pi^{m_d}_{n_d}@^2_{i_d} \right)
B.
\end{align}
\end{itemize}
\end{prp}

\noindent{\it Proof.} What the formula (\ref{3.2.4}) or (\ref{3.2.5}) did is  compressing (decompressing) each component by projections one by one. It is not difficult to see that the order of compressing (decompressing) components of the signal does not affect the result.

\hfill $\Box$

Next, we consider the matrix expressions of (\ref{3.2.4}) and (\ref{3.2.5}).
For $2$-dimensional signals, it is easy to get their matrix expressions.

\begin{cor}\label{c2.3.4} Assume $A\in {\cal M}_{m\times n}$.  It is compressed into $B\in {\cal M}_{s\times t}$ and decompressed back to $\hat{A}$ by projections. Then
\begin{itemize}
\item[(i)] The projection-based compression of $A$ into $B$ can be realized by
\begin{align}\label{3.2.6}
B:=\Pi^m_s A (\Pi^n_t)^T.
\end{align}
\item[(ii)] The projection-based decompression of $B$ into $\hat{A}$ can be realized by
\begin{align}\label{3.2.7}
\hat{A}:=\Pi^{s}_{m} B (\Pi^{t}_n)^T.
\end{align}
\end{itemize}
\end{cor}

When the order $d>2$ we need some preparations.

\begin{dfn}\label{d2.3.5} Assume a hypermatrix $A\in \R^{n_1\times \cdots \times n_d}$ as shown in (\ref{3.2.1}).
The vector form of $A$, denoted by $V_A$, is a column vector of dimension $n=\prod{i=1}^d n_i$, which has all elements of $A$ as its entries and arranged in the order that
$$
a_{i_1,i_2,\cdots,i_d}\prec a_{j_1,j_2,\cdots,j_d},
$$
if and only if, There exists a $1\leq k\leq d-1$, such that
$$
\begin{cases}
i_s=j_s,\quad s\leq k,\\
i_s<j_s,\quad s=k+1.
\end{cases}
$$
\end{dfn}

Denote by
$$
\begin{array}{l}
n^k:=\prod_{i=1}^kn_i,\\
m^k:=\prod_{i=1}^km_i, \quad k\in[1,d].
\end{array}
$$

\begin{prp}\label{p2.3.6} Let $A\in \R^{n_1\times \cdots \times n_d}$ be a signal with order $d$ ($d\geq 2$).
\begin{itemize}
\item[(i)] The compression (\ref{3.2.4}) can be realized by
\begin{align}\label{3.2.8}
V_B=\left[\ltimes_{i=1}^{d}(I_{n^{i-1}}\otimes \Pi^{n_i}_{m_i})\right]V_A.
\end{align}
\item[(ii)] The decompression (\ref{3.2.5}) can be realized by
\begin{align}\label{3.2.9}
V_{\hat{A}}=\left[\ltimes_{i=1}^{d}(I_{m^{i-1}}\otimes \Pi^{m_i}_{n_i})\right]V_B.
\end{align}
\end{itemize}
\end{prp}

\noindent{\it Proof.} We prove (\ref{3.2.8}) only. The proof of (\ref{3.2.9}) is the same.
Using the associativity of STP, the compression mapping can be realized by projections $\Pi^{n_k}_{m_k}$, $k=d,d_1,\cdots,1$ one by one.

First we consider the last projection $\Pi^{n_d}_{m_d}$.
Decompose $V_A$ into $n^{d-1}$ equal blocks as
$$
V_A=(A^T_{1},A^T_{2},\cdots, A^T_{n^{d-1}})^T,
$$
where any block, say block $\a$, can be expressed as
$$
A^T(\a)=(a_{i_1,\cdots,i_{d_1},1},  a_{i_1,\cdots,i_{d_1},2}, \cdots a_{i_1,\cdots,i_{d_1},n_d})^T.
$$
Then
$$
V^1_A:=
\begin{array}{l}
\left(I_{n^{d-1}}\otimes \Pi^{n_d}_{m_d}\right)V_A\\
=\begin{bmatrix}
\Pi^{n_d}_{m_d}A_1\\
\Pi^{n_d}_{m_d}A_2\\
\vdots\\
\Pi^{n_d}_{m_d}A_{d-1}\\
\end{bmatrix},
\end{array}
$$
which realizes
$$
V^1_A=\left(\Pi^{n_d}_{m_d}@^2_{i_d} \right)A.
$$

Second, converting the STP into conventional matrix product, we have
$$
V^2_A=\left(I_{n^{d-2}}\otimes \Pi^{n_{d-1}}_{m_{d-1}}\otimes I_{m_d}\right)V^1_A.
$$
Similar argument shows that
$$
V^2_A=\left(\Pi^{n_{d-1}}_{m_d}@^2_{i_{d-1}} \right)V^1_A.
$$
Continuing this procedure verifies (\ref{3.2.8}).

\hfill $\Box$

\subsection{Right PC and PD}

 In the aforementioned arguments about compression-decompression are based on the left project.
We may replace the left project by right project. That is,
Replace (\ref{2.2.1})  by
\begin{align}\label{3.3.1}
\langle x,y\rangle_{{\cal V}}:=\frac{1}{t} \langle (\J_{t/m}\otimes x),(\J_{t/n}\otimes y) \rangle.
\end{align}
Then the norm  (\ref{2.2.2}) and the distance  (\ref{2.2.3}) are also changed correspondingly.
The right projection is still formally defined by (\ref{2.3.2}), but where
$$
%\Pi^m_n=\frac{n}{t}\left(I_n\otimes \J^T_{t/n}\right)\left(I_m\otimes \J_{t/m}\right).
\Pi^m_n=\frac{n}{t}\left(\J^T_{t/n}\otimes I_n\right)\left(\J_{t/m}\otimes I_m\right).
$$
Using right projection,  all the compression-decompression based on left projection can be paralleled transformed to the one based on right projection. It seems that the right projection-based compression-decompression and the left projection-based
ones are not symmetrically related. Hence the  right projection-based compression-decompression is another approach. In certain case, say, for periodic signals,  it might be better than the left one.

The signal space for right system is
$$
\Omega=\R^{\infty}/\lra_{r}.
$$

\section{Compressed Sensing vs PC/PD}

\subsection{Compressed Sensing}

Compressed sensing (CS) is a newly appeared method for compression/decompression \cite{dua11}. The compressed sensing is a compression denoted by
\begin{align}\label{4.1.1}
y=Ax,
\end{align}
where $x\in \R^n$ is the signal, $y\in \R^m$ ($m<n$) is the measurement, and $A\in {\cal M}_{m\times n}$ is a sensing matrix.

The $\spark(A)$ is the largest number of columns of $A$ which are dependent. Then we have the following result.

\begin{prp}\label{p4.1.1} \cite{don03} Consider equation (\ref{4.1.1}). Assume
 $\spark(A)>2k$, then the equation has at most one solution $x\in \Sigma^n_k$.
 \end{prp}

 The coherence of $A$, denoted by  $\mu(A)$, is defined by
\begin{align}\label{4.1.2}
\mu(A)=\max_{1\leq i\neq j\leq n}\frac{<a_i,a_j>}{\|a_i\|\|a_j\|},
\end{align}
where, $a_i=\Col_i(A)$.
Then we have
\begin{prp}\label{p4.1.2} A sufficient condition for precisely recovering signal is \cite{dua11}
\begin{align}\label{4.1.3}
k<\frac{1}{2}\left(1+\frac{1}{\mu(A)}\right),
\end{align}
where $k$ is the number of nonzero entries of $x$.
\end{prp}

Propositions \ref{p4.1.1} and \ref{p4.1.2}  shown that  under certain condition CS approach is lossless, which shows that the CS broken the
Nyquist sampling theorem, which says that to avoid aliasing the sample frequency should be greater than twice of signal highest frequency \cite{ham77}.

Construct the sensing matrix is one of the fundamental issues in CS. Recently, the STP-CS has been proposed by  \cite{xie16}. Then it has received a quick development \cite{wan17,wen20}. The basic idea
is to replace the matrix-vector product in (\ref{4.1.1}) by MV-STP as
\begin{align}\label{4.1.4}
y=A_0\lvtimes x.
\end{align}

Note that (\ref{4.1.4}) is a little bit different from the original expression in  \cite{xie16}, where the MM-STP is used. Using MV-STP can remove the dimension restriction that $m|n$. As $m|n$ they are the same.

 The availability of STP-CS is based on the following proposition.

\begin{prp}\label{p1.4}\cite{xie16} Consider $A$ and $A\otimes I_s$, where $A\in {\cal M}_{m\times n}$, $m<n$, and $s\in \Z^+$.
\begin{itemize}
\item[(i)]
\begin{align}\label{4.1.5}
\spark(A\otimes I_s)=\spark(A).
\end{align}
\item[(ii)]
\begin{align}\label{4.1.6}
\mu(A\otimes I_s)=\mu(A).
\end{align}
\end{itemize}
\end{prp}

There are two obvious advantages of STP-CS: (i) It can reduce the storage of sensing matrix; (ii) it can be used for all $n$ dimensional signals.

\subsection{Various Topologies on $\R^{\infty}$}

Both the STP=CS and  the projection-based compression-decompression are based on STP. A comparison to reveal their difference is necessary.  It is obvious that no matter how small the compression rate is, the projection-based compression-decompression is not lossless. But  as we shown in last subsection that the STP=CS may cause lossless compression-decompression. Does it violate Theorem \ref{t3.1.1}? To answer this question, we need to make the topologies over $\R^{\infty}$ clearly.

Consider the set of sequences of countable real numbers, denoted by
\begin{align}\label{4.2.1}
{\cal V}^{\infty}:=\underbrace{\R\times \R\times \cdots\times \R\times \cdots}_{\infty}.
\end{align}
Then it is clear that ${\cal V}^{\infty}$ is a vector space. Moreover,
\begin{align}\label{4.2.2}
\R^{\infty}\subset {\cal V}^{\infty}
\end{align}
is its subspace.

We try to establish a topological structure over  ${\cal V}^{\infty}$, and then pose the subspace topology on $\R^{\infty}$.

In most existing signal compression-decompression, we merge $\R^{\infty}$ into  ${\cal V}^{\infty}$ in the following way:

\begin{dfn}\label{d4.2.1}
Assume $x\in \R^{\infty}$, specified by $x=(x_1,\cdots,x_n)\in \R^n$, then maps $x$ into ${\cal V}^{\infty}$ as
$$
\varphi (x)=(x_1,\cdots,x_n,0,0,\cdots)\in {\cal V}^{\infty}.
$$
Then we can pose a topology on  ${\cal V}^{\infty}$ by establishing a norm on it. Say,
\begin{itemize}
\item[(i)] $\ell_0$ pseudo-norm:
\begin{align}\label{4.2.3}
\|x\|_{0}:=\#(\mbox{nonzero component of}~ x).
\end{align}
The $\ell_0$ space
\begin{align}\label{4.2.4}
\ell_0:=\{x\in {\cal V}^{\infty}\;|\; \|x\|_{0}<\infty\}.
\end{align}
\item[(ii)]  $\ell_p$ norm ($p\geq 1$):
\begin{align}\label{4.2.5}
\|x\|_{p}:=\left[\dsum_{i=1}^{\infty}|x_i|^p\right]^{1/p}.
\end{align}
The $\ell_p$ space
\begin{align}\label{4.2.6}
\ell_p:=\{x\in {\cal V}^{\infty}\;|\; \|x\|_{p}<\infty\}.
\end{align}
\end{itemize}
\end{dfn}

The following result is obvious.

\begin{prp}\label{p4.2.2}
\begin{itemize}
\item[(i)] $\ell_0$ is a Fr\`{e}chet space; $\ell_p$, $p\geq 1$ are Banach spaces;   $\ell_2$ is a Hilbert space\footnote{We refer to any text book for functional analysis, say, \cite{tay80}, for the concepts and basic properties of these spaces.}.
\item[(ii)] Consider $x\in \R^{\infty}$ as an element $\varphi(x)\in {\cal V}^{\infty}$.
\begin{align}\label{4.2.7}
\R^{\infty}\subset \ell_p,\quad \forall p\in \Z^+.
\end{align}
\item[(iii)]
Using $\ell_0$ pseudo-norm, or $\ell_p$ norms, $p\geq 0$, the corresponding distances can be obtained as
$$
d_p(x,y):=\|x=y\|_{p},\quad x,y\in \ell_p,~p\in \Z^+.
$$
Then for each $p\in \Z^+$ the $\ell_p$ becomes a distance space. The topology deduced by $d_p$ is denoted by ${\cal T}_{\ell_p}$.
\item[(iv)] As a subspace of $\ell_p$, the inherent topology of $\R^{\infty}$ makes it a topological space, which is denoted by
$$
(\R^{\infty},{\cal T}_{\ell_p}),\quad p\in \Z^+.
$$
\end{itemize}
\end{prp}

A commonly used decompression method for CS is to find the $\ell_0$ optimal solution. While   solving $\ell_0$ optimal problem is an NP-hard problem, the   $\ell_1$ optimal solution is used to approximate the  $\ell_0$ one. And under certain
condition they coincide \cite{dua11}. Now it is clear that Theorem \ref{t3.1.1} says the projection-based compression-decompression is the best one is under ${\cal T}_d$ topology, while CS approach is under ${\cal T}_{\ell_0}$ and ${\cal T}_{\ell_1}$  topologies.

\section{Conclusion}

This paper proposes a technique for signal compression/decompression via projections between Euclidian spaces with different dimensions. It is proved that under the assumption of equivalence both encoding and decoding are optimal in the sense that the approximation errors are of least square. The encoding/dicoding formulas are presented for arbitrary dimensional signals with arbitrary preassigned compression rates.

This is only a theoretical result. The application examples for real signals are under preparation.

\end{document}